\documentclass[pre,twocolumn]{revtex4}
\usepackage{enumerate}
\usepackage{graphicx}
\begin{document}
\title{Holding strategies in a bus-route model}
\author{Sam A. Hill\email{drsamahill@gmail.com}}
\affiliation{Department of Physics, Southern Methodist University}
\date{September 1, 2007}
\begin{abstract}
A major source of delays in public transportation is the clustering instability, which causes late buses to become progressively later while the buses trailing it become progressively earlier.  In this paper, we study this instability and how to neutralize it using the common practices of holding and schedule slack.  Starting with an on-time route, we delay one or more buses at a single stop, and determine how these delays grow over time.  We compare the effects of two different types of holding on the stability of the system, and briefly investigate how our results change with the use of timepoints.
\end{abstract}
\maketitle

%======================================================================
\section{Introduction}
The typical bus or train system is intended to provide regular, periodic service, so that each stop has an arrival every $n$ minutes.  In practice, however, public transportation often suffers from maddening inconsistencies, which discourages potential passengers from depending on it for their daily commute.  Given the environmental, economic, and political impact of increased gasoline consumption, it is important for us to understand these delays and reduce their frequency, if possible.

One cause of these delays is an inherent instability\cite{Welding} in the dynamics of a bus route: when a single bus (bus A) is delayed, it has to pick up more passengers which delays it further, while the bus following it (bus B) has fewer passengers to pick up, and runs faster.  In many cases, bus A is unable to recover and, on busy routes, bus B may even catch up to bus A, forming a \emph{cluster}.  This is not always a problem: for an evening outbound route, where most passengers board at the beginning of the route, it does not matter that the service is irregular so long as the passengers are gotten to their destinations quickly.  However, in most circumstances, this instability results in longer, unpredictable passenger waits.

Bus dispatchers can counter this behavior with \emph{holding} and \emph{slack}.  Buses which are running too fast (such as bus B above) are \emph{held} at a stop so that they do not catch up to the bus in front.  In \emph{schedule-based holding}, buses are prevented from leaving a stop until a scheduled departure time, while in \emph{headway-based holding}, buses are prevented from leaving until the preceding bus is far enough away.  Most bus services implement holding only at a few \emph{time points} along the route, while light rail services, which typically stop at every station anyway, can implement holding at every stop.

If holding is only implemented when buses are running more quickly than usual, then it is only partially effective in preventing clusters: it may keep bus B from catching up to bus A, but it does not allow bus A to recover from its delay.  However, if most buses are held, then \emph{not} holding a delayed bus may be enough to allow it to recover.  This is done by introducing \emph{slack} into the schedule; that is, by allowing more time for buses to travel from stop to stop than is ordinarily necessary, so that the typical bus is held at every stop.

The clustering instability was first mentioned in the engineering literature by Welding\cite{Welding} in 1957; since then, it has been covered extensively in the engineering literature\cite{NewellPotts,Newell76,Engineering}, although not to the point of exhaustion.  A few physicists, inspired perhaps by the much larger field of traffic study\cite{Helbing}, have turned their attention to modelling bus routes\cite{OLoan,Krbalek,Nagatani,Nagatani-other,Hill}, using a number of methods (cellular automata, linear stability analysis, and chaos theory).

The engineering literature has primarily focused on the statistical properties of multiple delays along an entire bus route.  In this paper we take a more ``microscopic'' approach, by investigating the deterministic propagation of delays which occur at a single stop, on a bus route that is otherwise running on time.  Using a model which is similar to others in the literature\cite{NewellPotts,Nagatani,Hill}, we calculate the maximum delay that a bus can recover from (a quantity we call the ``buffer'') for a given amount of slack, and how this buffer varies when two or more consecutive buses are delayed at the same stop.  We show that headway-based holding is typically better in recovering from random delays, although schedule-based holding allows buses to recover more quickly when the delays are shorter.  We mostly assume that buses are held at every stop; however, we end by briefly showing the effect that timepoints have on the buffer of a single bus.

%======================================================================
\section{Model}

%----------------------------------------------------------------------
\begin{figure}[ht]\begin{center}
\includegraphics[width=3in]{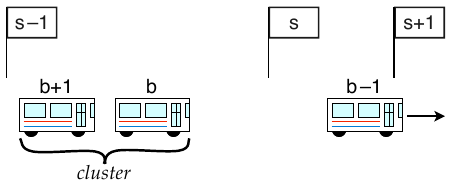}
\caption{\label{fig-indices}How buses and stops are indexed in this model.}
\end{center}\end{figure}
%----------------------------------------------------------------------
Consider a series of irregularly spaced bus stops, labelled with the index $s$, and a series of buses labelled with the index $b$.   Each bus visits each stop in increasing order, and each stop is visited by each bus in increasing order (Fig.~\ref{fig-indices}).  We define $t_{b,s}$ to be the time (in minutes) at which bus $b$ departs stop $s$.  Before bus $b$ can depart stop $s$, it must do three things:
\begin{enumerate}
\item depart stop $s-1$, at time $t_{b,s-1}$;
\item drive from stop $s-1$ to stop $s$, which takes time~$T_s$
; and
\item pick up passengers at stop $s$.
\end{enumerate}
As a simplification, we ignore the time it takes for passengers to get off the bus: this corresponds to the situation in which the number of alighting passengers is either very small, or in which passengers can alight through a rear door at the same time other passengers are boarding.

 The time it takes to pick up passengers at stop $s$ is equal to the time it takes each passenger to board (the \emph{unit boarding time}), multiplied by the number of passengers waiting at the stop.  This number of passengers is, in turn, equal to the wait since the last bus ($t_{b,s}-t_{b-1,s}$) divided by the time it takes each passenger to arrive (the \emph{interarrival time}, the reciprocal of the arrival frequency).  Thus, the time it takes to board passengers is
\begin{equation}
\hbox{boarding time}=\mu(t_{b,s}-t_{b-1,s}),
\end{equation}
with the \emph{passenger constant} $\mu>0$ defined as
\begin{equation}
\label{mu-definition}\mu={\hbox{unit boarding time}\over\hbox{interarrival time}}.
\end{equation}
We can now specify a recursion relation for $t_{b,s}$:
\begin{equation}
\label{dEvolution}t_{b,s}=t_{b,s-1}+T_s+\mu(t_{b,s}-t_{b-1,s}).
\end{equation}
This can be simplified with a change of variable \hbox{$\displaystyle t_{b,s}\to t_{b,s}+\sum_{i=0}^sT_{i}$}, which eliminates $T_s$ from the equation entirely.  If we then solve Eq.~\ref{dEvolution} for $t_{b,s}$ (and $T_s$ removed), we have
\begin{equation}
\label{mainequation}t_{b,s}=(1+\mu')t_{b,s-1}-\mu't_{b-1,s},
\end{equation}
where we define \begin{equation}
\label{muprime}\mu'={\mu\over1-\mu}.
\end{equation}

To implement \emph{holding}, we must define a \emph{schedule function} $S_{b,s}$, which specifies the time at which bus $b$ \emph{should} depart from stop $s$.  For simplicity, we assume a homogeneous schedule in which buses are evenly spaced:
\begin{equation}
\label{schedule}S_{b,s}=b\Delta+s(\mu\Delta+\sigma).
\end{equation}
The parameter $\Delta$ is the time between successive buses at a particular stop, $\mu\Delta$ is the time it takes a particular bus to travel from stop to stop under normal conditions, and $\sigma$ is the \emph{slack} built into the schedule\footnote{Note that $S_{b,s}$ would include an additional term $\sum_k T_k$, if we hadn't normalized $T_s$ out of our equations.}.  If there is no slack ($\sigma=0$), then $t_{b,s}=S_{b,s}$ is a solution to Eq.~\ref{mainequation}.

We implement \emph{holding} by specifying some \emph{earliest departure time} $t^{\min}_{b,s}$; if the bus is ready to leave before $t^{\min}$, it is held so that it leaves at $t^{\min}$.  We do this by rewriting Eq.~\ref{mainequation} as the conditional
\begin{equation}
\label{holding1}t_{b,s}=\max\cases{
	(1+\mu')t_{b,s-1}-\mu't_{b-1,s}\cr
	t^{\min}_{b,s}\cr}.
\end{equation}
Note that our model will hold buses at \emph{every} stop, rather than at the less frequent \emph{time points} seen in most bus systems.

It is convenient to work with the \emph{delay} of a bus at any given stop, rather than its departure time.  We define the \emph{unnormalized delay} of bus $b$ at stop $s$ to be
\begin{equation}
\label{lateness}\ell_{b,s}=t_{b,s}-S_{b,s},
\end{equation}
and the \emph{normalized delay} (or, simply, the \emph{delay}) to be
\begin{equation}
\label{normlate}d_{b,s}={\mu\over\sigma}\ell_{b,s}.
\end{equation}
Delay is measured relative to the schedule function Eq.~\ref{schedule}; early buses have $d<0$.
We rewrite Eq.~\ref{holding1} in terms of delay, to arrive at our main \emph{dynamic equation}
\begin{equation}
\label{basic}
d_{b,s}=\max\cases{
		(1+\mu')d_{b,s-1}-\mu'd_{b-1,s}-\mu'\cr
		cd_{b-1,s}\cr}
\end{equation}
where $cd_{b-1,s}$
is the \emph{minimum delay} of bus $b$ at stop $s$.  If $c=0$, then buses are never allowed to be early (and the delay $d_{b,s}$ is never negative), and we have schedule-based holding.  If $c=1$, then buses are never allowed to be earlier than the bus preceding them ($d_{b,s}\ge d_{b-1,s}$), and we have headway-based holding.  
When the second of the two conditions in Eq.~\ref{basic} is larger, we say that ``holding has been triggered''.

In the calculations that follow, it will be necessary to solve several recursion relations of the form 
\begin{equation}
\label{recursion}x_{s}=z x_{s-1}+Az^s+\gamma;
\end{equation}
this has the solution
\begin{equation}
\label{recursionsolution}x_s={\gamma\over 1-z}+\left(x_0-{\gamma\over1-z}\right)z^s+Asz^s.
\end{equation}

%======================================================================
\section{Solutions}
We now consider a case in which, following a series of on-time buses ($d_{b,s}=0$ for $s<0$ and $b<1$), one or more buses, starting with bus $b=1$, are delayed at stop $s=0$.

%----------------------------------------------------------------------
\subsection{The First Bus}\label{firstbussection}
\begin{figure}\begin{center}
\includegraphics[width=3in]{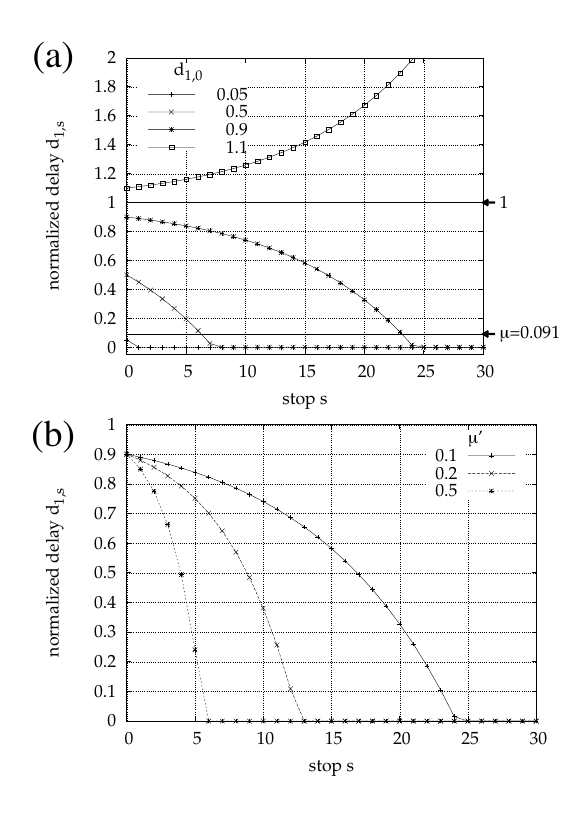}
\caption{\label{F-firstbus}The normalized delay of the first bus, calculated from Eq.~\ref{firstbus}.  In Fig.~\ref{F-firstbus}a, the passenger constant is fixed at $\mu'=0.1$, while the initial delay is varied: examples of unrecoverable, recovering, and instantly recovering buses are shown.  In Fig.~\ref{F-firstbus}b, the initial delay is fixed at $d_{1,0}=0.9$, while the passenger constant is varied. Notice that the bus recovers faster when the passenger constant is greater.}
\end{center}\end{figure}

Given our assumption that $d_{b,s}=0$ for $b<1$, the dynamic equation (Eq.~\ref{basic}) for the first bus is
\begin{equation}
\label{firstbus}d_{1,s+1}=\max\left[(1+\mu')d_{1,s}-\mu',0\right].
\end{equation}
If the bus is early or on-time at any stop $s_0$ ($d_{1,s_0}\le0$), it will be on-time from then on ($d_{1,s>s_0}=0$).
As long as $d_{1,s}>0$, however, Eq.~\ref{firstbus} has the solution (cf Eq.~\ref{recursionsolution})
\begin{equation}
\label{firstbussoln}d_{1,s}=1-(1+\mu')^s(1-d_{1,0})
\end{equation}
which grows exponentially towards positive or negative infinity, depending on the sign of $1-d_{1,0}$.  Thus there are two types of results (as shown in Fig.~\ref{F-firstbus}):
\begin{enumerate}[(a)]

\item If $d_{1,0}<1$, then the delay of the bus decreases until it is on time ($d_{1,s_0}\le0$), after which the bus remains on time.  We say that such a bus is \emph{recovering}.  The bus has fully recovered when
\begin{equation}
\label{s0}s>s_0\equiv{-\ln(1-d_{1,0})\over\ln(1+\mu')}.
\end{equation}
Note that bus routes with larger numbers of passengers (larger values of $\mu'$) recover more quickly (Fig.~\ref{F-firstbus}).  A special case of this is when $s_0<1$, in which case the bus \emph{recovers instantly}; this occurs when $d_{1,0}<\mu$.

\item If $d_{1,0}>1$, then the delay of the bus increases exponentially; we say that such a bus is \emph{unrecoverable}.  A single unrecoverable bus delays all following buses: either the bus behind it catches up to it (if $c=0$), forming a cluster, or else it also becomes exponentially late (if $c=1$); in either case, the homogeneous behavior has broken down irrevocably.  
\end{enumerate}
%----------------------------------------------------------------------
%----------------------------------------------------------------------

We define the \emph{buffer} $\beta$ of a bus to be the largest delay $d_{b,0}$ that it is able to recover from; the buffer of the first bus is thus $\beta_1=1$.  This is a normalized quantity based on the normalized delay $d_{b,s}$.  In terms of the \emph{unnormalized delay} $\ell_{b,s}$ (Eq.~\ref{lateness}), the bus recovers instantly when $\ell_{b,s}<\sigma$, because the bus need only eliminate part or all of its slack.  The bus's unnormalized buffer is $\sigma/\mu$: as Newell\cite{Newell76} explains it, if the slack per stop $\sigma$ is larger than the time $\mu\ell_{b,s}$ it takes to board the additional passengers due to the delay, then the bus will eventually recover.
%----------------------------------------------------------------------
\subsection{The Second Bus}
If the first bus is late, it will pick up some of the passengers originally destined for the second bus, and so the second bus will run faster.  Therefore, if the second bus starts out on-time, it will have no trouble remaining on-time, so long as the first bus is not unrecoverable.  It also stands to reason that if the second bus is delayed at stop $s=0$, then it will recover more quickly, and have a larger buffer $\beta$, than if the first bus were not also running late.

The dynamic equation (Eq.~\ref{basic}) for the second bus is
\begin{equation}
\label{secondbus}
d_{2,s}=\max\cases{
		(1+\mu')d_{2,s-1}-\mu'(d_{1,s}+1)\cr
		cd_{1,s}\cr},
\end{equation}
If the first bus is unrecoverable, then the second bus will be unrecoverable as well, so we need only consider the case where $d_{1,0}<1$.  If $d_{1,0}<\mu$, then the first bus recovers instantly, and the behavior of the second bus can be described using Section~\ref{firstbussection}.  Thus we only need consider the case where ${\mu'\over1+\mu'}<d_{1,0}<1$ (note that ${\mu'\over1+\mu'}=\mu$).  There are two cases to be considered:
 
\begin{enumerate}[(a)]
\item If the holding condition in Eq.~\ref{secondbus} is triggered at some stop $s-1$, then $d_{2,s-1}=cd_{1,s-1}$. Now as long as the first bus hasn't recovered, Eq.~\ref{firstbussoln} can be rewritten as
\begin{equation}
d_{1,s-1}=(d_{1,s}+\mu')/(1+\mu')
\end{equation}
and so 
\begin{equation}
\label{persistenthold2}d_{2,s}=\max\cases{
cd_{1,s}-\mu'd_{1,s}-\mu(1-c)\cr
cd_{1,s}}
.
\end{equation}
Clearly the upper term is smaller than the lower term, and so the holding condition is triggered at stop $s$ as well.  Therefore, once the holding condition has been triggered for bus 2, it will continue to be triggered from there on out.  If $c=0$, then the second bus recovers immediately; if $c=1$, then the second bus recovers along with the first bus.  In either case, however, once the holding condition is triggered, both buses recover.

\item If the second bus's holding condition is never triggered, and the first bus hasn't recovered yet (i.e. $s<s_0$), we can substitute Eq.~\ref{firstbussoln} into Eq.~\ref{secondbus} to get
\begin{equation}
\label{secondbus2}d_{2,s}=(1+\mu')d_{2,s-1}+\mu'(1+\mu')^s(1-d_{1,0})-2\mu'
,
\end{equation}
which, according to Eq.~\ref{recursionsolution}, has the solution
\begin{equation}
d_{2,s}=2+\left[\mu'(1-d_{1,0})s-(2-d_{2,0})\right](1+\mu')^s.
\end{equation}

Once $s>s_0$, $d_{1,s}=0$ and the second bus follows the ``first bus'' pattern; therefore, the second bus will ultimately recover only if $d_{2,s_0}<1$.
Since
\begin{equation}
d_{2,s_0}=2-{\mu'\ln(1-d_{1,0})\over\ln(1+\mu')}-{2-d_{2,0}\over 1-d_{1,0}},
\end{equation}
the inequality $d_{2,s_0}<1$ holds if $d_{2,0}<\beta_2$ where
\begin{equation}
\beta_2=1+d_{1,0}+{\mu'\over\ln(1+\mu')}(1-d_{1,0})\ln(1-d_{1,0}).
\end{equation}

\end{enumerate}

%----------------------------------------------------------------------
\begin{figure}\begin{center}
\includegraphics[width=3in]{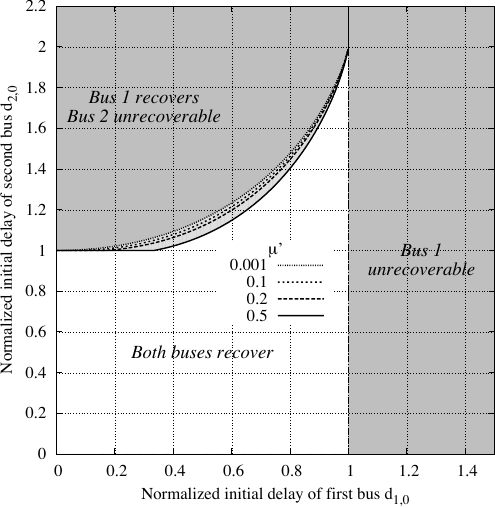}
\caption{\label{F-secondbus}A diagram showing the buffer of the second bus as a function of the first bus's initial delay $d_{1,0}$.  One can interpret this as a phase diagram: in the unshaded region, both buses recover, while in the shaded regions one or both buses are unrecoverable.  The buffer depends on the passenger constant $\mu'$, being smaller for larger passenger constants, but never going below 1.  This figure is independent of the holding strategy $c$.}
\end{center}\end{figure}
%----------------------------------------------------------------------

Figure~\ref{F-secondbus} shows a graph of the buffer of the second bus as a function of the first bus's initial delay.  The buffer's curve marks the transition between the case in which both buses recover (below), and the phase in which one or both buses are unrecoverable (above and to the right): we see that when the first bus is very late (but recovering), the second bus can recover from a much larger delay.  As the passenger constant $\mu'$ increases, the region where both buses recover becomes smaller; this is because, for larger passenger constants, the first bus recovers more quickly (as seen in Fig.~\ref{F-firstbus}) and so the second bus does not get as much benefit from the first bus's delay.

%----------------------------------------------------------------------

\subsection{The Third Bus}
%----------------------------------------------------------------------
\begin{figure}\begin{center}
\includegraphics[width=3in]{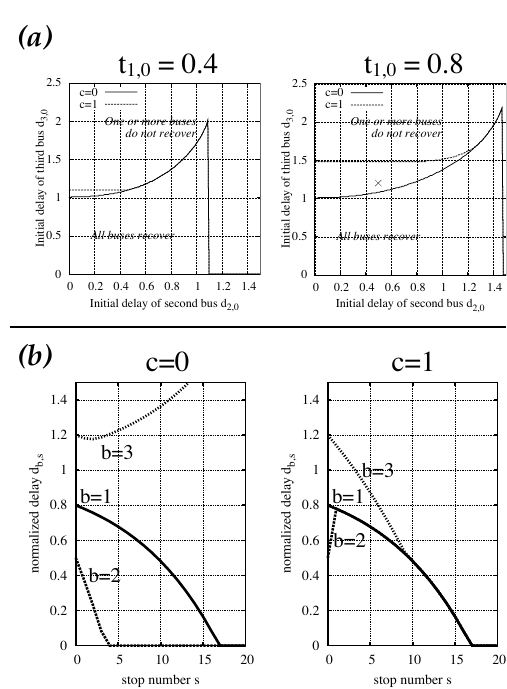}
\caption{\label{F-thirdbus}Figure~\ref{F-thirdbus}a shows the buffer of the third bus as a function of the delay of the second bus, and for two values of the delay of the first bus ($t_{1,0}=0.4$ and $t_{1,0}=0.8$); all three buses recover in the region underneath the corresponding buffer curve.  Clearly the buffer of the third bus depends on the holding strategy $c$. Figure~\ref{F-thirdbus}b shows how the three buses behave at the X marked in (a), with $t_{1,0}=0.8$.  For schedule-based holding ($c=0$), the third bus is unrecoverable, while for headway-based holding ($c=1$), all three buses recover.  The passenger constant is $\mu'=0.1$ throughout.}
\end{center}\end{figure}
%----------------------------------------------------------------------
Although the recovery of the first two buses is independent of the holding strategy $c$, the same is not true for subsequent delayed buses.  Fig.~\ref{F-thirdbus}a--b shows the buffer of the third bus ($b=3$) as it depends on the initial delay of the first two buses; this clearly depends on the holding strategy.  Fig.~\ref{F-thirdbus}c shows why this difference exists: if schedule-based holding is in place, then the second bus recovers quickly from a small initial delay, and the third bus loses the benefit of following a late bus. Note how, when $c=0$, the third bus's buffer approaches $1$ (the buffer of a bus following an undelayed bus) when $d_{2,0}$ approaches zero.  When $c=1$ and $d_{2,0}$ is small and $d_{1,0}$ is large, on the other hand, the second bus can only recover as quickly as the first bus, giving the third bus a larger buffer.  In a sense, the delay of a recovering bus is a \emph{resource} which makes the trailing buses more resistant to delay.

\subsection{Many Buses}

To expand our model to the many-bus limit in a manageable way, we consider the case where all buses are delayed by the same amount $\tau$ at stop $s=0$ (i.e. $d_{b,0}=\tau$ for all $b\ge1$); this might be caused by construction or traffic on one particular street.   If $\tau>\beta_1=1$, the first bus, and eventually all buses, will be unrecoverable; thus we only consider the case when $\tau<1$.  Any bus with $d_{b,0}<1$ will ultimately recover, so our concern is not whether this situation recovers, but how quickly it recovers.

\begin{figure}\begin{center}
\includegraphics[width=3in]{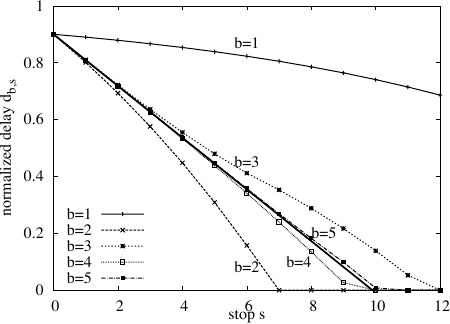}
\caption{\label{F-BulkLimit}The solution of Eq.~\ref{basic} when $d_{b,0}=\tau$ and $c=0$; buses alternate above and below the steady-state solution, shown as a solid black line here.}
\end{center}\end{figure}

For $c=0$ (Fig.~\ref{F-BulkLimit}), the solutions $d_{b,s}$ approach (in an alternating manner) the steady-state normalized solution $d_{b,s}=\tau-s\mu$ as $b$ increases.  This solution can be better understood by looking at the corresponding unnormalized solution $\ell_{b,s}={\sigma\over\mu}\tau-s\sigma$: buses make up time by eliminating slack from the schedule, becoming $\sigma$ minutes earlier at each stop until they have recovered.

%----------------------------------------------------------------------
%----------------------------------------------------------------------

For $c=1$, we can prove that 
$d_{b,s}=d_{1,s}$ for all $s$:
\begin{quotation}\noindent{\bf Proof:} by induction over $b$ and $s$.  The base case $b=1$ is automatic, while the base case $s=0$ is true because $d_{b,0}=d_{1,0}=\tau$.  Suppose that $d_{b',s'}=d_{1,s}$ for all $b'<b$ and $s'<s$.
Then Eq.~\ref{basic} becomes   
\begin{equation}
\label{bulk1}d_{b,s}=\max\cases{
(1+\mu')d_{1,s-1}-\mu'd_{1,s}-\mu'\cr
d_{1,s}\cr}.
\end{equation}
If $d_{1,s}=0$ and $d_{1,s-1}=0$, then $d_{b,s}=0$ and the hypothesis is satisfied.  If $d_{1,s-1}\ne 0$ but $d_{1,s}=0$, then $(1+\mu')d_{1,s-1}<\mu'$ (according to Eq.~\ref{firstbus}), which means that the first case in Eq.~\ref{bulk1} is negative, and so $d_{b,s}=0=d_{1,s}$ and the hypothesis is again satisfied.  If $d_{1,s}\ne0$ and $d_{1,s-1}\ne0$ then, according to Eq.~\ref{firstbus}, $(1+\mu')d_{1,s-1}=d_{1,s}+\mu'$, and so
\begin{equation}
d_{b,s}=\max\cases{
d_{1,s}+\mu'-\mu'd_{1,s}-\mu'\cr d_{1,s}\cr},
\end{equation}
and since $(1-\mu')d_{1,s}<d_{1,s}$, the second condition applies and $d_{b,s}=d_{1,s}$.  {\sc q.e.d.}
\end{quotation}
The steady-state solutions are thus 
\begin{equation}
\label{steadystate}
\lim_{b\to\infty} d_{b,s}=\cases{
\tau-s{\mu'\over1+\mu'}&$c=0$\cr
1-(1+\mu')^s(1-\tau)&$c=1$\cr
}.
\end{equation}
It can be shown that, as long as $d_{b,1}>0$ for both cases, the $c=0$ case in Eq.~\ref{steadystate} reaches zero faster than the $c=1$ case: this means that schedule-based holding allows the buses to recover more quickly than headway-based holding.
%======================================================================
\section{Timepoints}
So far we have considered the situation where holding occurs at every stop.  
This may be a realistic description of a light-rail system, where trains stop at every stop anyway.  However, a typical bus route may have a stop at every intersection, and a model which forces buses to stop and wait at each of these is unrealistic.  In reality, most bus systems designate a few stops as \emph{timepoints}, and only implement holding there.  To model this, we assume that every $N$th stop is a timepoint, and evaluate Eq.~\ref{basic} for $b=1$, with the stipulation that the holding condition only applies when $s\equiv 0\bmod N$.  As we vary $N$, we keep the amount of slack \emph{per stop}, which we call $\sigma$, constant; the amount of slack \emph{per timepoint} (that is, the amount of time a bus might actually have to wait while being held) is then $N\sigma$.  A computer simulation calculates the (dimensionless) buffer $\beta$ of a ``first bus'' by determining the initial delay $d_{1,0}$ for which the system moves from recovering ($d_{1,1000}<10$) and unrecoverable ($d_{1,1000}>10$: $10$ and $1000$ are both arbitrarily chosen).

%----------------------------------------------------------------------
\begin{figure}[ht]\begin{center}
\includegraphics[width=3in]{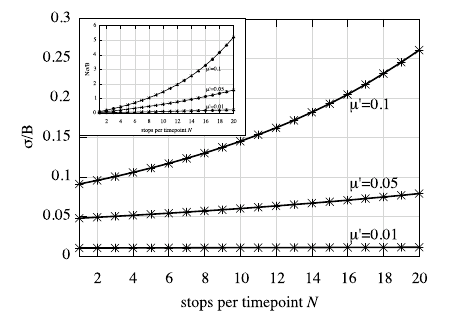}
\caption{\label{fig-timepoint}The amount of slack necessary per \emph{stop} for a given amount of buffer $B$, both measured in minutes, as the number of stops between timepoints increases.  The inset shows the amount of slack necessary per \emph{timepoint} (that is, $N\sigma/B$).  Data points were found using computer simulation.
}
\end{center}\end{figure}

If we let $B$ be the number of minutes corresponding to the dimensionless buffer $\beta$, then according to Eq.~\ref{normlate},
\begin{equation}
\label{def-B}B={\sigma\over\mu}\beta \Longrightarrow {\sigma\over B}={\beta\over\mu}.
\end{equation}
Figure~\ref{fig-timepoint} shows how this ratio $\sigma/B$ depends on the passenger constant $\mu'$ and the timepoint spacing $N$; for example, we find that for a busy bus route with $N=16$ and $\mu'=0.1$, the ratio $\sigma/B=0.21$, so the route must have 0.42~minutes' slack per stop, or a 6.7~minutes' slack per timepoint, to allow a bus to recover from a two-minute delay.  For small passenger constants, the amount of slack per stop remains fairly constant, as if the holding from each stop can be ``saved up'' until the next timepoint.  For larger passenger rates, however, holding becomes much less effective if the timepoints are spread too far apart.

%======================================================================
\section{Discussion}
When a single bus is delayed, it is able to recover if its delay is no larger than 
$\beta=1$ (or $B=\sigma/\mu$ minutes), which we call the bus's buffer; the buffer is even larger for buses that trail already-delayed buses.   

If the number of passengers increases (such as during rush hour), the buffer will be reduced unless steps are taken to keep it steady, either by increasing slack, or by decreasing the unit boarding time (Eq.~\ref{mu-definition}); this is part of the reason that bus routes are generally more unreliable during peak hours.  Since slack increases the time it takes to complete a route, one would like to find strategies that make it as small as possible: for example, it might be optimal to keep the slack proportional to the passenger arrival rate, keeping the buffer $\beta=\mu/\sigma$ constant, so that busier stops get more slack.  Alternatively, giving more slack to the stops immediately preceding or following the busier stops might be more advantageous; further study is required in this case.

When it comes to choosing a holding strategy, both schedule and headway-based holding have advantages depending on the circumstances.
When a series of buses experience the same delay at a given stop, schedule-based holding can allow the buses to recover more quickly as long as the delay is no larger than the buffer.  However, headway-based holding may be more appropriate in scenarios with larger, random delays, as a very delayed bus will have a greater probability of trailing a slightly delayed bus, and have a better opportunity to recover.

One interesting, possibly counterintuitive result of our analysis is that buses on busier routes recover more quickly from small delays.  The passenger constant seems to set the ``timescale'' of the route's behavior.

Our next step will be to more fully investigate the effect of timepoints on these results.  We have already shown that the necessary slack per stop increases dramatically when we only implement holding at every $N$th stop.  Future work will determine the exact nature of this relationship, and the effect that timepoints have on the behavior and buffers of subsequent buses.

The model presented here is a ``microscopic'' model, dealing on the effects of delays at a single stop; how these single-stop delays interact with one another over the course of an entire bus route is also something worth investigating.  Although engineers have spent a lot of time investigating this ``macroscopic'' regime, there may yet be basic truths which can be uncovered by a physicist's perspective.

We thank Dr. Peter Furth of Northeastern University for useful conversations and his engineering perspective.

\end{document}